\documentstyle[12pt,epsf]{article}

\def\baselinestretch{1.4}

\makeatletter
\@addtoreset{equation}{section}
\makeatother

\makeatletter
\def\citen{\protect\@p@citen}

\def\@p@citen#1{%
\edef\@tempa{\@ignspaftercomma,#1, \@end, }
\edef\@tempa{\expandafter\@ignendcommas\@tempa\@end}%
\if@filesw \immediate \write \@auxout {\string \citation {\@tempa}}\fi
\@tempcntb\m@ne    
\let\@h@ld\relax   
\let\@citea\@empty 
\let\@celt\over    
\def\@cite@list{}
\@for \@citeb:=\@tempa \do{\@make@cite@list}
\@tempcnta\m@ne    
\let\@celt\@compress@cite \@cite@list 
\@h@ld}

\begingroup \catcode`\_=8 
\gdef\@make@cite@list{%
 \expandafter\let \expandafter\@B@citeB
          \csname b@\@citeb\@extra@b@citeb \endcsname
 \ifx\@B@citeB\relax 
    \@citea {\bf ?}\let\@citea\citepunct
    \@warning {Citation `\@citeb' on page \thepage\space undefined}%
    \global\@namedef{b@\@citeb\@extra@b@citeb}{?}
 \else 
    \ifcat _\ifnum\z@<0\@B@citeB _\else A\fi 
       \@tempcnta\@B@citeB \relax
       \ifnum \@tempcnta>\@tempcntb 
          \edef\@cite@list{\@cite@list \@celt{\@B@citeB}}%
          \@tempcntb\@tempcnta
       \else 
          \edef\@cite@list{\expandafter\@sort@celt \@cite@list \@gobble @}%
       \fi
    \else 
       \@citea   
       \@B@citeB 
       \let\@citea\citepunct
 \fi\fi}
\endgroup 

\def\@compress@cite#1{
  \advance\@tempcnta\@ne 
  \ifnum #1=\@tempcnta   
     \ifx\@h@ld\relax    
        \edef \@h@ld{\@citea 
            #1
        }%
     \else               
        \edef \@h@ld{\hbox{--}\penalty\@highpenalty 
               #1
        }%
     \fi
  \else   
     \@h@ld \@citea 
     #1
     \let\@h@ld\relax
  \fi \@tempcnta#1\let\@citea\citepunct
}

\def\@sort@celt#1#2{\ifx \@celt #1
     \ifnum #2<\@tempcnta 
        \@celt{#2}%
        \expandafter\expandafter\expandafter\@sort@celt 
     \else 
        \@celt{\number\@tempcnta}\@celt{#2}
  \fi\fi}

\def\citepunct{,\penalty\@highpenalty\hskip.13emplus.1emminus.1em}%

\def\cite{\protect\@p@cite}

\def\@p@cite{\@ifnextchar [{\@tempswatrue\@citex}{\@tempswafalse\@citex[]}}

\def\@citex[#1]#2{\@cite{\@p@citen{#2}}{#1}}%

\def\@cite#1#2{\leavevmode\unskip
  \ifnum\lastpenalty=\z@ \penalty\@highpenalty \fi 
  \ [{\multiply\@highpenalty 3 
      #1\if@tempswa,\penalty\@highpenalty\ #2\fi 
    }]\spacefactor\@m}

\def\@ignspaftercomma#1, {\ifx\@end#1\@empty\else
   #1,\expandafter\@ignspaftercomma\fi}
\def\@ignendcommas,#1,\@end{#1}

\@ifundefined{@extra@b@citeb}{\def\@extra@b@citeb{}}{}

\makeatother

\def\BbbZ{{}\kern+1.6pt\hbox{$I$}\kern-7.5pt\hbox{$Z$}}
\def\figlab#1{\setbox0=\hbox{$#1)$}\wd0=0pt\ht0=0pt\dp0=0pt\box0}
\def\half{{\textstyle{1\over2}}}
    \let\p=\pi

\let\la=\label  
  
 \def\bd{\begin{document}} \def\ed{\end{document}}
\def\ds{\documentstyle} \let\fr=\frac \let\bl=\bigl \let\br=\bigr
\let\Br=\Bigr \let\Bl=\Bigl
\let\bm=\bibitem
\let\na=\nabla
\let\pa=\partial \let\ov=\overline
\newcommand{\be}{\begin{equation}}
\newcommand{\ee}{\end{equation}}
\def\ba{\begin{array}}
\def\ea{\end{array}}
\newcommand{\ho}[1]{$\, ^{#1}$}
\newcommand{\hoch}[1]{$\, ^{#1}$}
\newcommand{\bea}{\begin{eqnarray}}
\newcommand{\eea}{\end{eqnarray}}
\newcommand{\ra}{\rightarrow}
\newcommand{\lra}{\longrightarrow}
\newcommand{\Lra}{\Leftrightarrow}
\newcommand{\ap}{\alpha^\prime}
\newcommand{\bp}{\tilde \beta^\prime}
\newcommand{\tr}{{\rm tr} }
\newcommand{\Tr}{{\rm Tr} }
\newcommand{\NP}{Nucl. Phys. }

\newcommand{\tamphys}{\it Center for Theoretical Physics,
Department of Physics\\
Texas A\&M University, College Station, Texas 77843--4242}

\newcommand{\auth}{M. J. Duff, James T. Liu and R. Minasian}

\begin{document}

\hfill{CTP-TAMU-26/95}

\hfill{hep-th/9506126}

\vspace{24pt}

\begin{center}
{ \large {\bf
ELEVEN DIMENSIONAL ORIGIN OF STRING/STRING DUALITY: A ONE LOOP TEST%
\footnote{Research supported in part by NSF Grant PHY-9411543.}
}}

\vspace{36pt}

\auth

\vspace{10pt}

{\tamphys}

\vspace{44pt}

\underline{ABSTRACT}

\end{center}

Membrane/fivebrane duality in $D=11$ implies Type $IIA$ string/Type
$IIA$ fivebrane duality in $D=10$, which in turn implies Type $IIA$
string/heterotic string duality in $D=6$. To test the conjecture, we
reproduce the corrections to the $3$-form field equations of the $D=10$
Type $IIA$ string (a mixture of tree-level and one-loop effects)
starting from the Chern-Simons corrections to the $7$-form Bianchi
identities of the $D=11$ fivebrane (a purely tree-level effect). $K3$
compactification of the latter then yields the familiar gauge and
Lorentz Chern-Simons corrections to $3$-form Bianchi identities of the
heterotic string. We note that the absence of a dilaton in the $D=11$
theory allows us to fix both the gravitational constant and the
fivebrane tension in terms of the membrane tension.  We also comment on
an apparent conflict between fundamental and solitonic heterotic
strings and on the puzzle of a fivebrane origin of S-duality.

{\vfill\leftline{}\vfill}
\leftline{June 1995}
\newpage

\section{Introduction}
\la{Introduction}

With the arrival of the 1984 superstring revolution \cite{Green},
eleven-dimensional Kaluza Klein supergravity \cite{Pope} fell out of favor,
where it more or less remained until the recent observation by Witten
\cite{Witten} that $D=11$ supergravity corresponds to the strong coupling
limit of the $D=10$ Type $IIA$ superstring, coupled with the realization
that there is a web of interconnections between Type $IIA$ and all the
other known superstrings: Type $IIB$, heterotic $E_8\times E_8$, heterotic
$SO(32)$ and open $SO(32)$.  In particular, string/string duality
\cite{Luloop,Khurifour,Lublack,Minasian,Duffstrong,Khuristring,Duffclassical}
implies that the $D=10$ heterotic string compactified to $D=6$ on $T^4$ is
dual to the $D=10$ Type $IIA$ string compactified to $D=6$ on $K3$
\cite{Hull}.  Moreover, this automatically accounts for the conjectured
strong/weak coupling $S$-duality in $D=4$, $N=4$ supersymmetric theories,
since $S$-duality for one string is just target-space $T$-duality for the
other \cite{Duffstrong}.  In this paper we find further evidence for an
eleven-dimensional origin of string/string duality and hence for
$S$-duality.

$D=10$ string/fivebrane duality and $D=6$ string/string duality can
interchange the roles of spacetime and worldsheet loop expansions
\cite{Luloop}.  For example, tree-level Chern-Simons corrections to the
Bianchi identities in one theory may become one-loop Green-Schwarz
corrections to the field equations in the other.  In a series of papers
\cite{Lustrings,Luloop,Lubetween,Dixonputting,Izquierdo,Blum,Minasian,Vafa},
it has been argued that this provides a useful way of putting various
duality conjectures to the test.  In particular, we can compare quantum
spacetime effects in string theory with the $\sigma$-model anomalies
for the dual $p$-branes
\cite{Dixoncoupling,Dixonchern,Percacci,Bergshoeff1,Cederwall}
even though we do not yet know how to quantize the $p$-branes!  This is
the method we shall  employ in the present paper. We reproduce the
corrections to the $3$-form field equations of the $D=10$ Type $IIA$
string (a mixture of tree-level and one-loop effects) starting from the
Chern-Simons corrections to the $7$-form $\tilde K_7=*K_4$ Bianchi
identities of the $D=11$ fivebrane (a purely tree-level effect):
\begin{equation}
d{\tilde K}_7=-{1\over2}K_4{}^2 +(2\pi)^4{\tilde \beta}'{\tilde X}_8\ ,
\label{Bianchi7}
\end{equation}
where the fivebrane tension is given by $\tilde T_6=1/(2\pi)^3{\tilde\beta}'$
and where the $8$-form polynomial ${\tilde X}_8$ describes the
$d=6$ $\sigma$-model Lorentz anomaly of the $D=11$ fivebrane:
\begin{equation}
{\tilde X}_8=\frac{1}{(2\pi)^4}
\Bigl[-\frac{1}{768}(\tr R^2)^2+\frac{1}{192}\tr R^4\Bigr]
\ .
\label{anomaly}
\end{equation}
$K3$ compactification of (\ref{Bianchi7}) then yields the familiar gauge
and Lorentz Chern-Simons corrections to $3$-form Bianchi identities of
the heterotic string:
\begin{equation}
d{\tilde H}_3=\frac{1}{4}{\tilde \alpha}'(\tr F^2-\tr R^2)\ .
\label{Bianchi3}
\end{equation}

The present paper thus provides evidence not only for the importance of
eleven dimensions in string theory but also (in contrast to Witten's
paper) for the importance of supersymmetric extended objects with
$d=p+1>2$ worldvolume dimensions: the super $p$-branes%
\footnote{Super $p$-branes are reviewed in
\cite{Duffsuper,Townsendreview,Khuristring}}.

\section{Ten to eleven: it is not too late}
\label{sec:10eleven}

In fact it should have come as no surprise that string theory makes use
of eleven dimensions, as there were already tantalizing hints in this
direction:

$i$) In 1986, it was pointed out \cite{Nilsson1} that $D=11$
supergravity compactified on $K3\times T^{n-3}$  \cite{Nilsson2} and
the $D=10$ heterotic string compactified on $T^{n}$ \cite{Narain,Narain2}
have the same moduli spaces of vacua, namely
\begin{equation}
{\cal M}=\frac{SO(16+n,n)}{SO(16+n) \times SO(n)}\ .
\label{moduli}
\end{equation}
It was subsequently confirmed \cite{Seiberg,Aspinwall1}, in the context
of the $D=10$ Type $IIA$ theory compactified on $K3\times T^{n-4}$,
that this equivalence holds globally as well as locally.

$ii$) In 1987 the $D=11$ supermembrane was discovered
\cite{Bergshoeff2,Bergshoeff-annp}. It was then pointed out \cite{Howe}
that the $(d=2,D=10)$ Green-Schwarz action of the Type $IIA$
superstring follows by simultaneous worldvolume/spacetime dimensional
reduction of the $(d=3,D=11)$ Green-Schwarz action of the
supermembrane.

$iii$) In 1990, based on considerations of this $D=11$ supermembrane
which treats the dilaton  and moduli fields on the same footing, it was
conjectured \cite{Dufflu,Luduality} that discrete subgroups  of all the
old non-compact global symmetries of compactified supergravity
\cite{Scherk,Cremmer,Marcus,Dufffradkin} ({\it e.g.}~$SL(2,R)$,
$O(22,6)$, $O(24,8)$, $E_7$, $E_8$, $E_9$, $E_{10}$) should be promoted
to duality symmetries of either heterotic or Type $II$ superstrings.
The case for a target space $O(22,6;Z)$ ({\it $T$-duality}) had already
been made, of course \cite{Giveonreview}.  Stronger evidence for a
strong/weak coupling $SL(2,Z)$ ({\it $S$-duality}) in string theory was
subsequently provided in \cite{Font,Rey,Kalara,Sen2,Sen3,Schwarz1,Schwarz2,%
Binetruy,Khurifour,Sen4,Rahmfeld,Gauntlett,Khuristring}.  Stronger
evidence for their combination into an $O(24,8;Z)$ duality in heterotic
strings was provided in \cite{Rahmfeld,Duffclassical,Sen7,Rahmfeld2}
and stronger evidence for their combination into a discrete $E_7$ in
Type $II$ strings was provided in \cite{Hull}, where it was dubbed {\it
$U$-duality}.

$iv$) In 1991, the supermembrane was recovered as an elementary
solution of $D=11$ supergravity which preserves half of the spacetime
supersymmetry \cite{Stelle}.  ({\it Elementary} solutions are singular
and carry a Noether ``electric'' charge, in contrast to {\it solitons}
which are non-singular solutions of the source-free equations and carry
a topological ``magnetic'' charge.)  The preservation of half the
supersymmetries is intimately linked with the worldvolume kappa
symmetry. It followed by the same simultaneous dimensional reduction in
($ii$) above that the elementary Type $IIA$ string could be recovered
as a solution of Type $IIA$ supergravity. By truncation, one then
obtains the $N=1,D=10$ elementary string \cite{Dabholkar}.

$v$) In 1991, the elementary superfivebrane was recovered as a solution
of the dual formulation of $N=1,D=10$ supergravity which preserves half
of the spacetime supersymmetry \cite{Luelem}. It was then reinterpreted
\cite{Callan1,Callan2} as a non-singular soliton solution of the usual
formulation.  Moreover, it was pointed out that it also provides a
solution of both the Type $IIA$ and Type $IIB$ field equations
preserving half of the spacetime supersymmetry and therefore that there
exist both Type $IIA$ and Type $IIB$ superfivebranes.  This naturally
suggested a Type $II$ string/fivebrane duality in analogy with the
earlier heterotic string/fivebrane duality conjecture
\cite{Duffsuper,Strominger}. Although no Green-Schwarz action for the
$d=6$ worldvolumes is known, consideration of the soliton zero modes
means that the gauged fixed actions must be described by a chiral
antisymmetric tensor multiplet $(B^-_{\mu\nu},\lambda^I,\phi^{[IJ]})$
in the case of $IIA$ and a non-chiral vector multiplet
$(B_{\mu},\chi^I,A^I{}_J,\xi)$ in the case of $IIB$
\cite{Callan1,Callan2}.

$vi$) Also in 1991, black $p$-brane solutions of $D=10$ superstrings
were found \cite{Horowitz1} for $d=1$ ($IIA$ only), $d=2$ (Heterotic,
$IIA$ and $IIB$), $d=3$ ($IIA$ only), $d=4$ ($IIB$ only) $d=5$ ($IIA$
only), $d=6$ (Heterotic, $IIA$ and $IIB$) and $d=7$ ($IIA$ only).
Moreover, in the extreme mass=charge limit, they each preserve half of
the spacetime supersymmetry \cite{Luscan}.  Hence there exist all the
corresponding super $p$-branes, giving rise to $D=10$ particle/sixbrane,
membrane/fourbrane and self-dual threebrane duality conjectures in
addition to the existing string/fivebrane conjectures.  The soliton
zero modes are described by the supermultiplets listed in Table
(\ref{modes}).  Note that in contrast to the fivebranes, {\it both}
Type $IIA$ and Type $IIB$ string worldsheet supermultiplets are
non-chiral%
\footnote{This corrects an error in \cite{Luscan,Khuristring}}.
As such, they follow from $T^4$ compactification of the Type $IIA$
fivebrane worldvolume supermultiplets.

$vii$) In 1992, a fivebrane was discovered as a soliton of $D=11$
supergravity preserving half the spacetime supersymmetry \cite{Guven}.
Hence there exists a $D=11$ superfivebrane and it forms the subject of
the present paper. Once again, its covariant action is unknown but
consideration of the soliton zero modes means that the gauged fixed
action must be described by the same chiral antisymmetric tensor
multiplet in ($v$) above \cite{Gibbonstownsend,Townsendeleven,Khuristring}.
This naturally suggests a $D=11$ membrane/fivebrane duality.

$viii$) In 1993, it was recognized \cite{Luscan} that by dualizing a
vector into a scalar on the gauge-fixed $d=3$ worldvolume of the Type
$IIA$ supermembrane, one increases the number of worldvolume scalars
({\it i.e.}~transverse dimensions) from $7$ to $8$ and hence obtains the
corresponding worldvolume action of the $D=11$ supermembrane.  Thus the
$D=10$ Type $IIA$ theory contains a hidden $D=11$ Lorentz invariance!

$ix$) In 1994 \cite{Gibbons} and 1995 \cite{Horowitz2}, all the $D=10$
Type $IIA$ $p$-branes of ($vi$) above were related to either the $D=11$
supermembrane or the $D=11$ superfivebrane.

$x$) Also in 1994, the (extreme electric and magnetic black hole
\cite{Khurinew,Rahmfeld}) Bogomol'nyi spectrum necessary for the $E_7$
$U$-duality of the $D=10$ Type $IIA$ string compactified to $D=4$ on
$T^6$ was given an explanation in terms of the wrapping of either the
$D=11$ membrane or $D=11$ fivebrane around the extra dimensions
\cite{Hull}.

$xi$) In 1995, it was conjectured \cite{Townsendeleven} that the $D=10$
Type $IIA$ superstring should be identified with the $D=11$
supermembrane compactified on $S^1$, with the charged extreme black
holes of the former interpreted as the Kaluza-Klein modes of the
latter.

$xii$) Also in 1995, the conjectured duality of the $D=10$ heterotic
string compactified on $T^4$ and the $D=10$ Type $IIA$ string
compactified on $K3$ \cite{Hull,Witten}, combined with the above
conjecture implies that the $d=2$ worldsheet action of the $D=6$
($D=7$) heterotic string may be obtained by $K3$ compactification%
\footnote{The wrapping of the $D=10$ {\it heterotic} fivebrane
worldvolume around $K3$ to obtain a $D=6$ {\it heterotic} string
was considered in \cite{Minasian}.}
of the $d=6$ worldvolume action of the $D=10$ Type $IIA$ fivebrane
($D=11$ fivebrane) \cite{Townsendseven,Harvey}.  We shall shortly make
use of this result.

Following Witten's paper \cite{Witten} it was furthermore proposed
\cite{Bars} that the combination of perturbative and non-perturbative
states of the $D=10$ Type $IIA$ string could be assembled into $D=11$
supermultiplets.  It has even been claimed \cite{Aspinwall2} that both
the $E_8 \times E_8$ and $SO(32)$ heterotic strings in $D=10$ may be
obtained by compactifying the $D=11$ theory on $\Xi_1$ and $\Xi_2$
respectively, where $\Xi_1$ and $\Xi_2$ are one-dimensional structures
obtained by squashing $K3$!

\begin{table}
\halign{\indent #&\hfil # \hfil &\quad \hfil # \hfil &\quad \hfil # \hfil &
\quad \hfil# \hfil &\quad \hfil # \hfil&\quad #\hfil \cr
&$d = 7$ & Type IIA & $(A_{\mu}, \lambda, 3\phi)$ &  & $n = 1$&\cr
&$d = 6$ & Type IIA & $(B_{\mu\nu}^-, \lambda_R{}^I, \phi^{[IJ]})$ & $I = 1,
\ldots, 4$ & $(n_+, n_-) = (2,0)$&\cr
&   & Type IIB & $(B_{\mu}, \chi^I, A^I\,_J, \xi)$ & $I = 1, 2$ & $(n_+, n_-)
= (1,1)$&\cr
&        & Heterotic & $(\psi^a,\phi^{\alpha})$ &
$a = 1, \ldots, 60\hphantom{0}$ &\cr
\noalign{\vspace{-2mm}}
&&&& $\alpha=1,\ldots,120$ &$(n_+, n_-) = (1,0)$&\cr
&$d = 5$ & Type IIA & $(A_{\mu}, \lambda^I, \phi^{[IJ]\mid})$ & $I = 1, \ldots,
4$ & $n = 2$&\cr
&$d = 4$ & Type IIB & $(B_{\mu}, \chi^I, \phi^{[IJ]})$ & $I = 1, \ldots, 4$ &
$n = 4$&\cr
&$d = 3$ & Type IIA & $(\chi^I, \phi^I)$ & $I = 1, \ldots, 8$ & $n = 8$&\cr
&$d = 2$ & Type IIA & $(\chi_L{}^I, \phi_L{}^I), (\chi_R{}^I, \phi_R{}^I)$ &
$I = 1, \ldots, 8$ & $(n_+, n_-) = (8,8)$&\cr
&        & Type IIB & $(\chi_L{}^I, \phi_L{}^I), (\chi_R{}^I, \phi_R{}^I)$ &
$I = 1, \ldots, 8$ & $(n_+, n_-) = (8,8)$&\cr
&        & Heterotic & $(0, \phi_L{}^M), (\chi_R{}^I, \phi_R{}^I)$ & $M = 1,
\ldots, 24$ &\cr
\noalign{\vspace{-2mm}}
&        &           &                                              & $I=1,
\ldots, 8$ & $(n_+, n_-) = (8,0)$&\cr}

\def\baselinestretch{1}
\caption{Gauge-fixed $D=10$ theories on the worldvolume, corresponding
to the zero modes of the soliton, are described by the above
supermultiplets and worldvolume supersymmetries.  The $D=11$ membrane
and fivebrane supermultiplets are the same as Type $IIA$ in $D=10$.}
\label{modes}
\end{table}

\section{$D=11$ membrane/fivebrane duality}
\la{eleven}

We begin with the bosonic sector of the $d=3$ worldvolume of the $D=11$
supermembrane:
\begin{eqnarray}
S_3&=&T_3\int d^3\xi\biggl[-{1\over2}\sqrt{-\gamma}\gamma^{ij}
\partial_i X^M\partial_j X^N G_{MN}(X) +{1\over2}\sqrt{-\gamma}\nonumber\\
&&\qquad\qquad
-{1\over3!}\epsilon^{ijk}\partial_i X^M\partial_j X^N\partial_k X^P
C_{MNP}(X)\biggr]\ ,
\label{membrane}
\end{eqnarray}
where $T_3$ is the membrane tension, $\xi^i$ ($i=1,2,3$) are the
worldvolume coordinates, $\gamma^{ij}$ is the worldvolume metric and
$X^M(\xi)$ are the spacetime coordinates $(M=0,1,\ldots,10)$.  Kappa
symmetry \cite{Bergshoeff2,Bergshoeff-annp} then demands that the
background metric $G_{MN}$ and background 3-form potential $C_{MNP}$
obey the classical field equations of $D=11$ supergravity, whose
bosonic action is
\begin{equation}
I_{11}=\frac{1}{2\kappa_{11}{}^2}\int d^{11}x\sqrt{-G}
\left[R_G-\frac{1}{2\cdot4!}K_{\scriptscriptstyle MNPQ}^2\right]
-\frac{1}{12\kappa_{11}{}^2} \int C_3\wedge K_4 \wedge K_4 \ ,
\label{supergravity11}
\end{equation}
where $K_4=dC_3$ is the 4-form field strength. In particular, $K_4$
obeys the field equation
\begin{equation}
d*K_4=-{1\over2}K_4{}^2
\label{equation4}
\end{equation}
and the Bianchi identity
\begin{equation}
dK_4=0\ .
\label{Bianchi4}
\end{equation}
While there are two dimensionful parameters, the membrane tension $T_3$
and the eleven-dimensional gravitational constant $\kappa_{11}$, they are
in fact not independent.  To see this, we note from (\ref{membrane}) that
$C_3$ has period $2\pi/T_3$ so that $K_4$ is quantized according to
\begin{equation}
\int K_4={2\pi n\over T_3}\qquad n={\rm integer}\ .
\label{eq:kquant}
\end{equation}
Consistency of such $C_3$ periods with the spacetime action,
(\ref{supergravity11}), gives the relation
\begin{equation}
{(2\pi)^2\over\kappa_{11}{}^2T_3{}^3}\in 4\BbbZ\ .
\label{eq:k11t3}
\end{equation}

The $D=11$ classical field equations admit as a soliton a dual superfivebrane
\cite{Guven,Lublack} whose worldvolume action is unknown, but which couples
to the dual field strength $\tilde K_7=*K_4$.  The fivebrane tension
${\tilde T}_6$ is given by the Dirac quantization rule \cite{Lublack}
\begin{equation}
2\kappa_{11}{}^2 T_3 {\tilde T}_6 =2\pi n \qquad n={\rm integer}\ .
\label{Dirac11}
\end{equation}
Using (\ref{eq:k11t3}), this may also be written as
\begin{equation}
\pi {\tilde T_6\over T_3{}^2}\in \BbbZ\ ,
\label{eq:newdirac}
\end{equation}
which we will find useful below.  Although Dirac quantization rules of
the type (\ref{Dirac11}) appear for other $p$-branes and their duals in
lower dimensions \cite{Lublack}, it is the absence of a dilaton in the
$D=11$ theory that allows us to fix both the gravitational constant and
the dual tension in terms of the fundamental tension.

{}From (\ref{equation4}), the fivebrane Bianchi identity reads
\begin{equation}
d\tilde K_7=-{1\over2}K_4{}^2\ .
\end{equation}
However, such a Bianchi identity will in general require gravitational
Chern-Simons corrections arising from a sigma-model anomaly on the
fivebrane worldvolume \cite{Dixoncoupling,Dixonchern,Percacci,Bergshoeff1,%
Cederwall,Dixonputting,Minasian}:
\begin{equation}
d\tilde K_7=-{1\over2}K_4{}^2 + (2\pi)^4{\tilde \beta}'{\tilde X}_8\ ,
\label{Bianchi7q}
\end{equation}
where ${\tilde \beta}'$ is related to the fivebrane tension by
$T_6=1/(2\pi)^3{\tilde \beta}'$ and where the $8$-form polynomial
${\tilde X}_8$, quartic in the gravitational curvature $R$,
describes the $d=6$ $\sigma$-model Lorentz anomaly of the $D=11$
fivebrane.  Although the covariant fivebrane action is unknown, we know
from section \ref{sec:10eleven} that the gauge fixed theory is
described by the chiral antisymmetric tensor multiplet $(B^-_{\mu\nu},
\lambda^I, \phi^{[IJ]})$, and it is a straightforward matter to read
off the anomaly polynomial from the literature.  See, for example
\cite{Alvarez,Ginsparg}.  The contribution from the anti self-dual tensor is
\begin{equation}
{\tilde X}_B={1\over(2\pi)^4}{1\over5760}
\Bigl[-10(\tr R^2)^2+28\,\tr R^4\Bigr]
\end{equation}
and the contribution from the four left-handed (symplectic) Majorana-Weyl
fermions is
\begin{equation}
{\tilde X}_\lambda={1\over(2\pi)^4}{1\over5760}
\Bigl[{10\over4}(\tr R^2)^2+2\,\tr R^4\Bigr]\ .
\end{equation}
Hence ${\tilde X}_8$ takes the form quoted in the introduction:
\begin{equation}
{\tilde X}_8={1\over(2\pi)^4}
\Bigl[-{1\over768}(\tr R^2)^2+{1\over192}\tr R^4\Bigr]\ .
\la{X}
\end{equation}
Thus membrane/fivebrane duality predicts a spacetime correction to the
$D=11$ supermembrane action
\begin{equation}
I_{11}({\rm Lorentz})=T_3\int C_3\wedge
{1\over(2\pi)^4}\Bigl[-{1\over768}(\tr R^2)^2+{1\over192}\tr R^4\Bigr] \ .
\end{equation}
Unfortunately, since the correct quantization of the supermembrane is
unknown, this prediction is difficult to check.  However, by
simultaneous dimensional reduction \cite{Howe} of $(d=3,D=11)$ to
$(d=2,D=10)$ on $S^1$, this prediction translates into a corresponding
prediction for the Type $IIA$ string:
\begin{equation}
I_{10}({\rm Lorentz})=T_2\int B_2\wedge
{1\over(2\pi)^4}\Bigl[-{1\over768}(\tr R^2)^2+{1\over192}\tr R^4\Bigr] \ ,
\end{equation}
where $B_2$ is the string $2$-form, $T_2$ is the string tension,
$T_2=1/{2\pi\alpha'}$, related to the membrane tension by
\begin{equation}
T_2=2\pi RT_3\ ,
\label{T}
\end{equation}
where $R$ is the $S^1$ radius.

As a consistency check we can compare this prediction with
previous results found by explicit string one-loop calculations. These
have been done in two ways: either by computing directly in $D=10$ the
one-loop amplitude involving four gravitons and one $B_2$
\cite{Lerche,lnsw,swanomlet,swanomnp},
or by compactifying to $D=2$ on an $8$-manifold $M$ and computing the
$B_2$ one-point function \cite{Vafa}.  We indeed find agreement. In
particular, we note that
\begin{equation}
\tilde X_8={1\over6}[2Y_8^{NS,R}-Y_8^{R,R}]\ ,
\end{equation}
where
\begin{eqnarray}
Y_8^{NS,R}&=&{1\over{(2\pi)^4}}{1\over2880}
\Bigl[-{25\over4}(\tr R^2)^2+31\,\tr R^4\Bigr]
\nonumber\\
Y_8^{R,R}&=&{1\over{(2\pi)^4}}{1\over2880}
\Bigl[10 (\tr R^2)^2-28\,\tr R^4\Bigr]\ .
\end{eqnarray}
Upon compactification to $D=2$, we arrive at
\begin{eqnarray}
n_{NS,R}&=&\int_MY_8^{NS,R}\nonumber\\
n_{R,R}&=&\int_MY_8^{R,R}\ ,
\end{eqnarray}
where in the (NS,R) sector $n_{NS,R}$ computes the index of the Dirac
operator coupled to the tangent bundle on $M$ and in the (R,R) sector
$n_{R,R}$ computes the index of the Dirac operator coupled to the spin
bundle on $M$. We also find agreement with the well-known tree-level
terms
\begin{equation}
\frac{1}{2\kappa_{10}{}^2}\int {1\over2}B_2 \wedge K_4 \wedge K_4 \ ,
\end{equation}
where
\begin{equation}
\kappa_{11}{}^2=2\pi R\kappa_{10}{}^2\ .
\label{kappa}
\end{equation}
Thus using $D=11$ membrane/fivebrane duality we have correctly
reproduced the corrections to the $B_2$ field equations of the $D=10$
Type $IIA$ string (a mixture of tree-level and string one-loop effects)
starting from the Chern-Simons corrections to the Bianchi identities of
the $D=11$ superfivebrane (a purely tree-level effect). It is now
instructive to derive this same result from $D=10$ string/fivebrane
duality.

\section{$D=10$ Type $IIA$ string/fivebrane duality}
\la{ten}

To see how a double worldvolume/spacetime compactification of the
$D=11$ supermembrane theory on $S^1$ leads to the Type $IIA$ string in
$D=10$ \cite{Howe}, let us denote all $(d=3,D=11)$ quantities by a hat
and all $(d=2,D=10)$ quantities without.  We then make a ten-one split
of the spacetime coordinates
\be
{\hat X}^{\hat M}=(X^M,Y)\qquad M=0,1,\ldots,9
\ee
and a two-one split of the worldvolume coordinates
\begin{equation}
{\hat \xi}^{\hat i}= (\xi^i,\rho)\qquad i=1,2
\end{equation}
in order to make the partial gauge choice
\be
\rho=Y\ ,
\ee
which identifies the eleventh dimension of spacetime with the third
dimension of the worldvolume. The dimensional reduction is then
effected by taking $Y$ to be the coordinate on a circle of radius $R$
and discarding all but the zero modes.  In practice, this means taking
the background fields ${\hat G}_{{\hat M}{\hat N}}$ and ${\hat
C}_{{\hat M}{\hat N}{\hat P}}$ to be independent of $Y$.  The string
backgrounds of dilaton $\Phi$, string $\sigma$-model metric $G_{MN}$,
$1$-form $A_M$, $2$-form $B_{MN}$ and $3$-form $C_{MNP}$ are given by%
\footnote{The choice of dilaton prefactor, $e^{-\Phi/3}$, is dictated by the
requirement that $G_{MN}$ be the $D=10$ string $\sigma$-model metric.  To
obtain the $D=10$ fivebrane $\sigma$-model metric, the prefactor is unity
because the reduction is then spacetime only and not simultaneous
worldvolume/spacetime.  This explains the remarkable ``coincidence''
\cite{Lublack} between $\hat G_{MN}$ and the fivebrane $\sigma$-model
metric.}
\begin{eqnarray}
{\hat G}_{MN}&=& e^{-\Phi/3}\left(
\begin{array}{cc}
G_{MN}+e^\Phi A_MA_N&e^{\Phi}A_M\\
e^{\Phi}A_N&e^{\Phi}
\end{array}
\right)\nonumber\\
{\hat C}_{MNP}&=&C_{MNP}\nonumber\\
{\hat C}_{MNY}&=&B_{MN}\ .
\end{eqnarray}

The actions (\ref{membrane}) and (\ref{supergravity11}) now reduce to
\begin{eqnarray}
S_2&=&T_2\int d^2\xi\biggl[-{1\over2}\sqrt{-\gamma}\gamma^{ij}
\partial_i X^M\partial_j X^N G_{MN}(X) \nonumber\\
&&\qquad\qquad
-{1\over2!}\epsilon^{ij}\partial_i X^M\partial_j X^N B_{MN}(X)+\cdots\biggr]
\label{string}
\end{eqnarray}
and
\begin{eqnarray}
I_{10}&=&\frac{1}{2\kappa_{10}{}^2}\int d^{10}x\sqrt{-G}e^{-\Phi} \left[
R_G+(\partial_{\scriptscriptstyle M}\Phi)^2
-\frac{1}{2\cdot3!}H_{\scriptscriptstyle MNP}^2
-\frac{1}{2\cdot2!}e^\Phi F_{\scriptscriptstyle MN}^2
-\frac{1}{2\cdot4!}e^\Phi J_{\scriptscriptstyle MNPQ}^2 \right]\nonumber\\
&&-{1\over2\kappa_{10}{}^2}\int{1\over2}K_4\wedge K_4\wedge B_2\ ,
\label{eq:supergravity10}
\end{eqnarray}
where the field strengths are given by $J_4=K_4+A_1H_3$, $H_3=dB_2$ and
$F_2=dA_1$.  Let us now furthermore consider a simple spacetime
compactification of the fivebrane theory on the same $S^1$ to obtain
the Type $IIA$ fivebrane in $D=10$.  From (\ref{Bianchi4}) and
(\ref{Bianchi7q}), the field equations and Bianchi identities for the
field strengths $J_4$, $H_3$, $F_2$ and their duals $\tilde J_6=*J_4$,
$\tilde H_7=e^{-\Phi}*H_3$, $\tilde F_8=*F_2$ now read
\begin{eqnarray}
dJ_4&=&{F_2H_3}\hphantom{0}\qquad d\tilde J_6=H_3J_4\\
dH_3&=&{0}\hphantom{F_2H_3}\qquad d\tilde H_7=-{1\over2}J_4{}^2+F_2\tilde J_6
+(2\pi)^4 {\tilde \beta}'{\tilde X}_8\\
dF_2&=&{0}\hphantom{F_2H_3}\qquad d\tilde F_8=-H_3\tilde J_6\ .
\end{eqnarray}
Of course, the Lorentz corrections to the Bianchi identity for $\tilde
H_7$ could have been derived directly from the Type $IIA$ fivebrane in
$D=10$ since its worldvolume is described by the same antisymmetric
tensor supermultiplet.  Note that of all the Type $IIA$ $p$-branes in Table
(\ref{modes}), only the fivebrane supermultiplet is chiral, so only the
$\tilde H_7$ Bianchi identity acquires corrections.

{}From (\ref{Dirac11}), (\ref{T}) and (\ref{kappa}), or from first
principles of string/fivebrane duality \cite{Luremarks}, the Dirac
quantization rule for $n=1$ is now
\be
2\kappa_{10}{}^2=(2\pi)^5\alpha'{\tilde \beta}'\ .
\ee
So from either $D=10$ string/fivebrane duality or from compactification
of $D=11$ membrane/fivebrane duality, the $B_2$ field equation with its
string one-loop correction is
\begin{equation}
d(e^{-\Phi}*H_3)=-{1\over2}J_4{}^2+F_2*J_4
+{2\kappa_{10}{}^2\over2\pi\alpha'}{\tilde X}_8\ ,
\end{equation}
which once again agrees with explicit string one-loop calculations
\cite{Lerche,Vafa}.

\section{$D=7$ string/membrane duality}
\label{sec:seven}

Simultaneous worldvolume/spacetime compactification of the $D=11$
fivebrane on $K3$ gives a heterotic string in $D=7$
\cite{Townsendseven,Harvey}.  The five worldvolume scalars produce
$(5_L,5_R)$ worldsheet scalars, the four worldvolume fermions produce
$(0_L,8_R)$ worldsheet fermions and the worldvolume self-dual $3$-form
produces $(19_L,3_R)$ worldsheet scalars, which together constitute the
field content of the heterotic string.  We may thus derive the Bianchi
identity for this string starting from the fivebrane Bianchi identity,
(\ref{Bianchi7}):
\begin{equation}
d{\tilde K}_7=-{1\over2}K_4{}^2 + (2\pi)^4{\tilde \beta}'{\tilde X}_8\ .
\label{eq:sevenB}
\end{equation}
We begin by performing a seven-four split of the eleven-dimensional
coordinates
\begin{equation}
X^M=(x^\mu,y^i)\qquad \mu=0,1,\ldots,6\ ;\ i=7,8,9,10
\end{equation}
so that the original set of ten-dimensional fields $\{{\cal A}_n\}$ may be
decomposed in a basis of harmonic $p$-forms on $K3$:
\begin{equation}
{\cal A}_n(X) = \sum {\cal A}_{n-p}(x)\omega_p(y)\ .
\end{equation}
In particular, we expand $C_3$ as
\begin{equation}
C_3(X)=C_3(x)+{1\over2T_3}\sum C_1^I(x)\omega_2^I(y)\ ,
\end{equation}
where $\omega_2^I$, $I=1,\ldots,22$ are an integral basis of $b_2$ harmonic
two-forms on $K3$.  We have chosen a normalization where the seven-dimensional
$U(1)$ field strengths $K_2^I=dC_1^I$ are coupled to even
charges
\begin{equation}
\int K_2^I\in 4\pi \BbbZ\ ,
\label{eq:u1norm}
\end{equation}
which follows from the eleven-dimensional quantization condition,
(\ref{eq:kquant}).

Following \cite{Minasian}, let us define the dual (heterotic) string tension
${\tilde T}_2=1/2\pi{\tilde \alpha}'$ by
\begin{equation}
\frac{1}{2\pi{\tilde \alpha}'}=\frac{1}{(2\pi)^3{\tilde \beta}'}V\ ,
\end{equation}
where $V$ is the volume of $K3$, and the dual string $3$-form
${\tilde H}_3$ by
\begin{equation}
\frac{1}{2\pi{\tilde \alpha}'}{\tilde H}_3
=\frac{1}{(2\pi)^3{\tilde \beta}'}\int_{K3}{\tilde K}_7\ ,
\end{equation}
so that $\tilde H_3$ satisfies the conventional quantization condition
\begin{equation}
\int\tilde H_3={4\pi^2}n{\tilde \alpha}'\ ,
\end{equation}
which follows from the underlying $\tilde K_7$ quantization.
The dual string Lorentz anomaly polynomial, ${\tilde X}_4$, is given by
\begin{eqnarray}
{\tilde X}_4=\int_{K3}\tilde X_8&=&{1\over(2\pi)^4}\int_{K3} \left[
-{1\over768}(\tr R^2+\tr R_0^2)^2+{1\over192}(\tr R^4+\tr R_0^4)
\right]\nonumber\\
&=&{1\over(2\pi)^2}{1\over192}\tr R^2 p_1(K3)\nonumber\\
&=&-{1\over(2\pi)^2}{1\over4}\tr R^2\ ,
\end{eqnarray}
where $p_1(K3)$ is the Pontryagin number of $K3$
\begin{equation}
p_1(K3)=-{1\over8\pi^2}\int_{K3}\tr R_0^2=-48\ .
\end{equation}
We may now integrate (\ref{eq:sevenB}) over $K3$, using the Dirac
quantization rule, (\ref{eq:newdirac}), to find
\begin{equation}
d\tilde H_3=-{\tilde\alpha'\over4}\left[K_2^IK_2^Jd_{IJ}+\tr R^2\right]\ ,
\label{eq:d7bi}
\end{equation}
where $d_{IJ}$ is the intersection matrix on $K3$, given by
\begin{equation}
d_{IJ} = \int_{K3}\omega_2^I\wedge\omega_2^J
\end{equation}
and has $b_2^+=3$ positive and $b_2^-=19$ negative eigenvalues.  Therefore
we see that this form of the Bianchi identity corresponds to a $D=7$
toroidal compactification of a heterotic string at a generic point on
the Narain lattice \cite{Narain,Narain2}.  Thus we have reproduced
{\it exactly} the $D=7$ Bianchi identity of the heterotic string,
starting from a $D=11$ fivebrane!

\section{$D=6$ string/string duality}
\la{six}
\label{sec:k3compact}

Further compactification of (\ref{eq:d7bi}) on $S^1$ clearly yields the
six-dimensional Bianchi identity with two additional $U(1)$ fields coming
from $S^1$, giving $\tr F^2$ with signature $(4,20)$.  Alternatively,
this may be obtained from $K3$ compactification of the $D=10$ fivebrane,
with Bianchi identity
\begin{equation}
d\tilde H_7=-{1\over2}J_4{}^2+F_2\tilde J_6+(2\pi)^4\tilde\beta'\tilde X_8\ .
\label{eq:bianchi7}
\end{equation}
Although in this section we focus just on this identity, we present the
compactification of the complete bosonic $D=10$ Type $IIA$ action,
(\ref{eq:supergravity10}), in the Appendix.

The reduction from ten dimensions is similar to that from eleven.  There
is one subtlety, however, which is that $J_4$ is the $D=10$ gauge invariant
combination, $J_4=K_4+A_1H_3$.  Compactifying (\ref{eq:bianchi7}) to six
dimensions on $K3$, we may identify 22 $U(1)$ fields coming from the
reduction of $J_4$ and one each coming from $F_2$ and $\tilde J_6$.
Normalizing these 24 six-dimensional $U(1)$ fields according to
(\ref{eq:u1norm}), we obtain
\begin{equation}
d\tilde H_3=-{\tilde\alpha'\over4}\left[J_2^IJ_2^Jd_{IJ}-2F_2\tilde J_2
-16\pi^2\tilde X_4\right]\ ,
\label{eq:dh3int}
\end{equation}
where $J_2^I=dC_2^I+A_1db^I$ and $J_4=dC_3+A_1H_3$.  The 22 scalars $b^I$
are torsion moduli of $K3$.  While we may be tempted to identify these
two-forms with $U(1)$ field strengths, this would not be correct since
$dJ_2^I=F_2db^I\ne0$ and $d\tilde J_2=J_2^Idb^Jd_{IJ}\ne0$.  Thus the
actual field strengths must be shifted according to
\begin{eqnarray}
\hat K_2^I&=&J_2^I-F_2b^I\nonumber\\
\hat J_2&=&\tilde J_2 - J_2^Ib^Jd_{IJ}
+{\textstyle{1\over2}}F_2b^Ib^Jd_{IJ}\ ,
\label{eq:shiftj2}
\end{eqnarray}
so that $d\hat K_2^I=d\hat J_2=0$.  Inverting these definitions and
inserting them into (\ref{eq:dh3int}) gives finally
\begin{equation}
d\tilde H_3=-{\tilde\alpha'\over4}\left[\hat K_2^I\hat K_2^Jd_{IJ}
-2F_2\hat J_2 +\tr R^2\right]\ .
\end{equation}

In order to compare this result with the toroidally compactified
heterotic string, it is useful to group the $U(1)$ field-strengths
into a 24-dimensional vector
\begin{equation}
{\cal F}_2=[F_2,\ \hat J_2,\ \hat K_2^I]^T\ ,
\end{equation}
in which case the $D=6$ Bianchi identity now reads
\begin{equation}
d\tilde H_3=-{\tilde\alpha'\over4}\left[{\cal F}^T L{\cal F}+\tr R^2\right]
\ ,
\end{equation}
where the matrix $L=[(-\sigma^1)\oplus d_{IJ}]$ has 4 positive and 20 negative
eigenvalues.  This is in perfect agreement with the reduction of the $D=7$
result, (\ref{eq:d7bi}), and corresponds to a Narain compactification
on $\Gamma_{4,20}$.

Note that the heterotic string tension $1/2{\pi}{\tilde \alpha}'$ and
the Type $IIA$ string tension $1/2{\pi}\alpha'$ are related by the
Dirac quantization rule \cite{Lublack,Minasian}
\begin{equation}
2\kappa_6{}^2=(2\pi)^3n\alpha'{\tilde \alpha}'\ ,
\end{equation}
where $\kappa_6{}^2=\kappa_{10}{}^2/V$ is the $D=6$ gravitational
constant.  Some string theorists, while happy to endorse string/string
duality, eschew the {\it soliton} interpretation. It is perhaps worth
emphasizing, therefore, that without such an interpretation with its
Dirac quantization rule, there is no way to relate the two string
tensions.

\section{Elementary versus solitonic heterotic strings}
\la{controversy}

Our success in correctly reproducing the {\it fundamental} heterotic
string $\sigma$-model anomaly polynomial
\be
{X}_4=\frac{1}{4}\frac{1}{(2\pi)^2}(\tr R^2-\tr F^2)\ ,
\la{4}
\ee
by treating the string as a ($K3$ compactified fivebrane) {\it
soliton}, now permits a re-evaluation of a previous controversy
concerning fundamental \cite{Gross} versus solitonic
\cite{Luremarks,Lustrings,Khuristring} heterotic strings.  In an
earlier one loop test of $D=10$ heterotic string/heterotic fivebrane
duality \cite{Dixonputting}, $X_4$ was obtained by the following logic:
the $d=2$ gravitational anomaly for complex fermions in a
representation $\cal{R}$ of the gauge group is \cite{Alvarez,Ginsparg}
\be
I_4=\frac{1}{2}\frac{1}{(2\pi)^2}(\frac{r}{24}\tr R^2 - \tr_{\cal R}F^2)\ ,
\ee
where $r$ is the dimensionality of the representation and $R$ is the
{\it two-dimensional} curvature.  Since the $SO(32)$ heterotic string
has $32$ left-moving gauge Majorana fermions (or, if we bosonize, $16$
chiral scalars) and $8$ physical right-moving spacetime Majorana
fermions, Dixon, Duff and Plefka \cite{Dixonputting} set ${\cal R}$ to
be the fundamental representation and put $r=32-8=24$ to obtain
$X_4=I_4/2$, on the understanding that $R$ is now to be interpreted as
the pull-back of the {\it spacetime} curvature.  Exactly the same logic
was used in \cite{Dixonputting} in obtaining the heterotic%
\footnote{Note that the heterotic string $X_4$, the heterotic fivebrane
$\tilde X_8$ and the Type $IIA$ fivebrane $\tilde X_8$ are the only
non-vanishing anomaly polynomials, since from Table (\ref{modes}), these are
the only theories with chiral supermultiplets.}
fivebrane ${\tilde X_8}$
\be
{\tilde X}_8 = \frac{1}{(2\p)^4}\,
 \Bl [ \frac{1}{24} \,\tr F^4 - \frac{1}{192}\,
\tr F^2 \, \tr R^2 + \frac{1}{768}\,(\tr R^2)^2 + \frac{1}{192}\, \tr R^4
\Br ]
\ee
and in sections \ref{eleven} and \ref{ten} above in obtaining the Type
$IIA$ fivebrane ${\tilde X_8}$ of (\ref{X}).  This logic was however
criticized by Izquierdo and Townsend \cite{Izquierdo} and also by Blum
and Harvey \cite{Blum}.  They emphasize the difference between the {\it
gravitational anomaly} (which vanishes for the {\it fundamental}
heterotic string \cite{Gross}) involving the {\it two-dimensional}
curvature and the {\it $\sigma$-model} anomaly (which is given by $X_4$
\cite{Hullwitten}) involving the pull-back of the {\it spacetime}
curvature.  Moreover, they go on to point out that the $32$ left-moving
gauge Majorana fermions (or $16$ chiral scalars) of the fundamental
heterotic string do not couple at all to the spin connections of this
latter curvature.  They conclude that the equivalence between $X_4$ and
$I_4/2$ is a ``curious fact'' with no physical significance. They would
thus be forced to conclude that the derivation of the Type $IIA$ string
field equations presented in the present paper is also a gigantic
coincidence!

An attempt to make sense of all this was made by Blum and Harvey.  They
observed that the zero modes of {\it solitonic} strings (and
fivebranes) necessarily couple to the spacetime spin connections
because they inherit this coupling from the spacetime fields from which
they are constructed.  For these objects, therefore, they would agree
that the logic of Dixon, Duff and Plefka (and, by inference, the logic
of the present paper) is correct.  But they went on to speculate that
although {\it fundamental} and {\it solitonic} heterotic strings may
both exist, they are {\it not} to be identified!  Recent developments
in string/string duality
\cite{Hull,Duffstrong,Witten,Harvey,Senssd,Townsendseven}, however,
have convinced many physicists that the fundamental heterotic string is
a soliton after all and so it seems we must look for an alternative
explanation.

The correct way to resolve the apparent conflict is, we believe, rather
mundane.  The solitonic string and $p$-brane solitons are invariably
presented in a {\it physical gauge} where one identifies $d$ of the $D$
spacetime dimensions with the $d=p+1$ dimensions of the $p$-brane
worldvolume.  As discussed in \cite{Dixonputting}, this is best seen in
the Green-Schwarz formalism, which is in fact the only formalism
available for $d>2$. In such a physical gauge (which is only
well-defined for vanishing {\it worldvolume} gravitational anomaly) the
worldvolume curvatures and pulled-back spacetime curvatures are mixed
up.  So, in this sense, the gauge fermions do couple to the spacetime
curvature after all.

\section{Fivebrane origin of S-duality?
\newline
Discard worldvolume Kaluza-Klein modes?}

In a recent paper \cite{Duffstrong}, it was explained how $S$-duality
in $D=4$ follows as a consequence of $D=6$ string/string duality:
$S$-duality for one theory is just $T$-duality for the other. Since we
have presented evidence in this paper that Type $IIA$ string/heterotic
string duality in $D=6$ follows as a consequence of Type $IIA$
string/Type $IIA$ fivebrane duality in $D=10$, which in turn follows
from membrane/fivebrane duality in $D=11$, it seems natural to expect a
fivebrane origin of $S$-duality.  (Indeed, a fivebrane explanation for
$S$-duality has already been proposed by Schwarz and Sen \cite{Schwarz1}
and by Binetruy \cite{Binetruy}, although they considered a $T^6$
compactification of the {\it heterotic} fivebrane rather than a $K3
\times T^2$ compactification of the {\it Type $IIA$ fivebrane}.)

The explanation of \cite{Duffstrong} relied on the observation that the
roles of the axion/dilaton fields $S$ and the modulus fields $T$ trade
places in going from the fundamental string to the dual string. It was
proved that, for a dual string compactified from $D=6$ to $D=4$ on
$T^2$, $SL(2,Z)_S$ is a symmetry that interchanges the roles of the
dual string worldsheet Bianchi identities and the field equations for
the internal coordinates $y^m$ $(m=4,5)$. However, in unpublished work
along the lines of \cite{Dufflu,Luduality},  Duff, Schwarz and Sen
tried and failed to prove that, for a fivebrane compactified from
$D=10$ to $D=6$, $SL(2,Z)_S$ is a symmetry that interchanges the roles
of the fivebrane worldvolume Bianchi identities and the field equations
for the internal coordinates $y^m$ $(m=4,5,6,7,8,9)$. A similar negative
result was reported by Percacci and Sezgin \cite{Percacci2}.

Another way to state the problem is in terms of massive worldvolume
Kaluza-Klein modes.  In the double dimensional reduction of the $D=10$
fivebrane to $D=6$ heterotic string considered in section \ref{six}, we
obtained the heterotic string worldsheet multiplet of $24$ left-moving
scalars,  $8$ right moving scalars and $8$ chiral fermions as the {\it
massless modes} of a Kaluza-Klein compactification on $K3$. Taken in
isolation, these massless modes on the dual string worldsheet will
display the usual $T$-duality when the string is compactified from
$D=6$ to $D=4$ and hence the fundamental string will display the
desired $S$-duality.  However, no-one has yet
succeeded in showing that this $T$-duality survives when the {\it
massive} Kaluza-Klein modes on the fivebrane worldvolume are included.
Since these modes are just what distinguishes a string
$X^M(\tau,\sigma)$ from a fivebrane $X^M(\tau,\sigma,\rho^i)$
$(i=1,2,3,4)$, this was precisely the reason in \cite{Duffstrong} for
preferring a $D=6$ string/string duality explanation for $SL(2,Z)$ over
a $D=10$ string/fivebrane duality explanation. (Another reason, of
course, is that the quantization of strings is understood, but that of
fivebranes is not!)  The same question about whether or not to discard
massive worldvolume Kaluza-Klein modes also arises in going from the
membrane in $D=11$ to the Type $IIA$ string in $D=10$.  For the moment
therefore, this inability to provide a fivebrane origin for $SL(2,Z)$
remains the Achilles heel of the super $p$-brane programme%
\footnote{Another unexplained phenomenon, even in pure string theory,
is the conjectured $SL(2,Z)$ duality of the $D=10$ Type $IIB$ string
\cite{Hull}, which gives rise to $U$-duality in $D=4$. In this
connection, it is perhaps worth noting from Table (1) that the
gauged-fixed worldvolume of the self-dual Type $IIB$ superthreebrane is
described by the $d=4,n=4$ Maxwell supermultiplet \cite{Luthree}.  Now
$d=4,n=4$ abelian gauge theories are expected to display an $SL(2,Z)$
duality.  See \cite{Wittendual,Verlinde} for a recent discussion. Could
this be the origin of the $SL(2,Z)$ of the Type $IIB$ string which
follows from a $T^2$ compactification of the threebrane?  Note
moreover, that the threebrane supermultiplet itself follows from $T^2$
compactification of either the Type $IIA$ or Type $IIB$ fivebrane
supermultiplet.  Compactifications of such $d=6$ self-dual
antisymmetric tensors have, in fact, recently been invoked precisely in
the context of $S$-duality in abelian gauge theories \cite{Verlinde}.
Of course, the gauged-fixed action for the superthreebrane is
presumably not simply the Maxwell action but some non-linear (possibly
Born-Infeld \cite{Luthree}) version.  Nevertheless, $S$-duality might
still hold \cite{Gibbonsnew}.}.

\section{Web of interconnections}
\la{web}

We have discussed membrane and fivebranes in $D=11$, heterotic strings
and Type $II$ fivebranes in $D=10$, heterotic strings and membranes in
$D=7$, heterotic and Type $II$ strings in $D=6$ and how they are
related by various compactifications.  This somewhat bewildering mesh
of interconnections is summarized in Fig.~(1$a$).  There are two types
of dimensional reduction to consider: lines sloping down left to right
represent spacetime reduction $(d,D)\rightarrow(d,D-k)$ and lines
sloping down right to left represent simultaneous spacetime/worldsheet
reduction $(d,D)\rightarrow(d-k,D-k)$.  The worldsheet reductions may
be checked against Table (1).  Note that the simultaneous reduction on
$\Xi$ of the $D=11$ membrane to yield the $D=10$ heterotic string is
still somewhat speculative \cite{Aspinwall2}, but we have included it
since it nicely completes the diagram.

According to Townsend \cite{Townsendseven}, a similar picture may be
drawn relating the Type $IIA$ string and heterotic fivebrane, which we
show in Fig.~(1$b$), where we have once again speculated on a spacetime
reduction on $\Xi$ of the $D=11$ fivebrane to yield the $D=10$
heterotic fivebrane.  However, one must now explain how $T^3$ (or
$T^4$) compactification of the $(120,120)$ degrees of freedom of the
gauge-fixed $D=10$ heterotic fivebrane \cite{Strominger} can yield only
the $(8,8)$ of the $D=7$ membrane (or the $(8_L,8_L),(8_R,8_R)$ of the $D=6$
Type $IIA$ string). Townsend has given arguments to support this claim.
There are more interrelationships one can illustrate by including
horizontal lines representing worldsheet reduction only,
$(d,D)\to(d-k,D)$, some of which are shown in Figs.~(2$a$,$b$).

Note that these diagrams describe theories related by compactification
and so relate weak coupling to weak coupling and strong to strong.  In
Fig.~(3), we have superimposed Figs.~(1$a$) and ($1b$) to indicate how
the various theories are also related by {\it duality} (denoted by the
dotted horizontal lines) which relates weak coupling to strong.  We
believe that these interrelationships, which have in particular enabled
us to deduce supermembrane effects in agreement with explicit string
one-loop calculations, strengthen the claim
that eleven dimensions and supermembranes have a part to play in string
theory: a triumph of {\it diversification} over {\it unification}
\cite{Dyson}.

\section{Acknowledgements}

We are grateful to Chris Hull, Joachim Rahmfeld, Paul Townsend and Edward
Witten for conversations.

\appendix
\section{Reduction of the $D=10$ Type $IIA$ model on $K3$}
In section \ref{sec:k3compact} we presented the reduction of the fivebrane
Bianchi identity on $K3$.  For completeness, we present the reduction of
the bosonic part of the $D=10$ Type $IIA$ supergravity action,
(\ref{eq:supergravity10}), which we write here in a form notation:
\begin{eqnarray}
I_{10}&=&\frac{1}{2\kappa_{10}{}^2}\int d^{10}x\sqrt{-G}e^{-\Phi} \left[
R_G+(\partial_M\Phi)^2\right]\nonumber\\
&&+{1\over4\kappa_{10}{}^2}\int \left[F_2\wedge *F_2
+e^{-\Phi}H_3\wedge *H_3 + J_4\wedge *J_4 - K_4\wedge K_4\wedge B_2\right]\ ,
\label{eq:sugra10}
\end{eqnarray}
where the ten-dimensional bosonic fields are the metric $G$, dilaton $\Phi$
and the 1-, 2- and 3-form fields $A_1$, $B_2$ and $C_3$.  Eleven-dimensional
$K_4$ quantization, (\ref{eq:kquant}), as well as the usual Kaluza-Klein
condition for $F_2$, give rise to the ten-dimensional conditions
\begin{eqnarray}
\int K_4&=&{4\pi^2nR\over T_2}\nonumber\\
\int H_3&=&{2\pi n\over T_2}\nonumber\\
\int F_2&=&2\pi nR\ .
\end{eqnarray}

Following the decomposition of the fields in section \ref{sec:seven}, we
write
\begin{eqnarray}
A_1(X)&=&{R\over2}A_1(x)\nonumber\\
B_2(X)&=&B_2(x)+{2\pi\over T_2}\sum b^I(x)\omega_2^I(y)\nonumber\\
C_3(X)&=&{R\over2}C_3(x)+{\pi R\over T_2}\sum C_1^I(x)\omega_2^I(y)\ ,
\end{eqnarray}
in which case the four-form $J_4$ is given by
\begin{equation}
J_4(X)={R\over2}[K_4(x)+A_1(x)H_3(x)]+{\pi R\over T_2}
\sum[K_2^I(x)+A_1(x)db^I(x)]\omega_2^I(y)\ .
\end{equation}
The constants are chosen so the six-dimensional $U(1)$ fields will
be coupled to even charges
\begin{equation}
\int {\cal F}_2\in 4\pi \BbbZ\ .
\end{equation}

For $K3$, with Betti numbers $b_0=1$, $b_1=0$, $b_2^+=3$ and $b_2^-=19$,
we may choose an integral basis of harmonic two-forms, $\omega_2^I$ with
intersection matrix
\begin{equation}
d_{IJ}=\int_{K3}\omega_2^I\wedge\omega_2^J\ .
\end{equation}
Since taking a Hodge dual of $\omega_2^I$ on $K3$ gives another harmonic
two-form, we may expand the dual in terms of the original basis
\begin{equation}
\hat*\omega_2^I=\omega_2^JH^J{}_I\ ,
\end{equation}
where we use $\hat *$ to denote Hodge duals on $K3$.  In this case, we find
\begin{equation}
\int_{K3}\omega_2^I\wedge\hat*\omega_2^J=d_{IK}H^K{}_J\ .
\end{equation}
The matrix $H^I{}_J$ depends on the metric on $K3$, {\it i.e.}~the
$b_2^+\cdot b_2^-=57$ $K3$ moduli.  Because of the fact
that $\hat*\hat*=1$, $H^I{}_J$ satisfies the properties \cite{Harvey}
\begin{eqnarray}
H^I{}_JH^J{}_K&=&\delta^I{}_K\nonumber\\
d_{IJ}H^J{}_K&=&d_{KJ}H^J{}_I\ ,
\end{eqnarray}
so that
\begin{equation}
H^J{}_Id_{JK}H^K{}_L=d_{IL}
\end{equation}
and hence is an element of $SO(3,19)/SO(3)\times SO(19)$.

Using these properties of $K3$, we may compactify the second line of
(\ref{eq:sugra10}) to obtain
\begin{eqnarray}
I_6&=&{1\over2\kappa_6{}^2}\int\biggl[
\half e^{-\phi}H_3\wedge *H_3
+\half e^{-\phi}e^\rho db^I\wedge *db^Jd_{IK}H^K{}_J\nonumber\\
&&\qquad+{\tilde\alpha'\over4}\Bigl(
e^{-\rho}F_2\wedge *F_2+e^{-\rho}J_4\wedge *J_4
+J_2^I\wedge *J_2^J d_{IK}H^K{}_J\nonumber\\
&&\qquad\qquad
- K_2^I\wedge K_2^J\wedge B_2\,d_{IJ}
- 2K_4\wedge K_2^Ib^Jd_{IJ}\Bigr)\biggr]\ .
\end{eqnarray}
The six-dimensional dilaton is given by $\phi=\Phi+\rho$ where $\Phi$
is the ten-dimensional dilaton and $\rho$ is the breathing mode of $K3$:
\begin{equation}
e^{-\rho}={1\over V}\int_{K3}\hat*1\ .
\end{equation}
In order to make contact with the compactified heterotic string, we
wish to dualize the four-form $J_4$.  Note, however, that since
$d(e^{-\rho}{\hat*}J_4)=J_2^Idb^Jd_{IJ}$, the proper expression for
dualizing $J_4$ is given by (\ref{eq:shiftj2}).  Performing such a step
and rewriting $J_2^I$ as well, we finally arrive at
\begin{eqnarray}
I_6&=&{1\over2\kappa_6{}^2}\int\biggl[
\half e^{-\phi}H_3\wedge *H_3+\half e^{-\phi}e^\rho
db^I\wedge *db^Jd_{IK}H^K{}_J\nonumber\\
&&\qquad+{\tilde\alpha'\over4}\Bigl(e^{-\rho}F_2\wedge *F_2
+(\hat K_2^I+F_2b^I)\wedge *(\hat K_2^J+F_2b^J)d_{IK}H^K{}_J\nonumber\\
&&\qquad\qquad +e^\rho (\hat J_2+\hat K_2^Ib^Jd_{IJ}+\half F_2b^Ib^Jd_{IJ})
\wedge * (\hat J_2+\hat K_2^Kb^Ld_{KL}+\half F_2b^Kb^Ld_{KL})\nonumber\\
&&\qquad\qquad-
(\hat K_2^I\wedge \hat K_2^Jd_{IJ}-2F_2\wedge \hat J_2)\wedge B_2
\Bigr)\biggr]\ .
\end{eqnarray}

This expression can be brought into a $SO(4,20)/SO(4)\times SO(20)$
invariant form.  As in section \ref{sec:k3compact}, we group the $U(1)$
field strengths into the 24 component vector
\begin{equation}
{\cal F}_2=[F_2,\ \hat J_2,\ \hat K_2^I]^T\ ,
\end{equation}
which allows us to rewrite the bosonic lagrangian as
\begin{eqnarray}
I_6&=&{1\over2\kappa_6{}^2}\int d\,^6x\sqrt{-G}e^{-\phi}
\left(R+(\partial_\mu\phi)^2-{1\over2\cdot3!}H_{\mu\nu\lambda}^2
+{1\over8}\Tr [\partial_\mu ML\partial_\mu ML]\right)\nonumber\\
&&+{1\over2\kappa_6{}^2}\int {\tilde\alpha'\over4}
\left({\cal F}_2{}^T(LML)\wedge*{\cal F}_2
-{\cal F}_2{}^T\wedge L {\cal F}_2 \wedge B_2\right)\ .
\end{eqnarray}
The matrix $L$ is given by
\begin{equation}
L=\left[\matrix{-\sigma^1&0\cr0&d_{IJ}\cr}\right]\ ,
\end{equation}
where $\sigma^1=\left({0\ 1\atop 1\ 0} \right)$.  The matrix $M$ contains the
$1+57+22=80$ moduli of $K3$ with torsion, broken up in terms of $e^\rho$,
$H^I{}_J$ and $b^I$ respectively:
\begin{equation}
M=\left[\matrix{e^\rho&-{1\over2}e^\rho(b^Ib^Jd_{IJ})&e^\rho b^I\cr
-{1\over2}e^\rho(b^Ib^Jd_{IJ})&e^{-\rho}+b^Ib^Jd_{IK}H^K{}_J
+{1\over4}e^\rho(b^Ib^Jd_{IJ})^2&-b^KH^I{}_K-{1\over2}e^\rho
b^I(b^Kb^Ld_{KL})\cr
e^\rho b^J&-b^KH^J{}_K-{1\over2}e^\rho b^J(b^Kb^Ld_{KL})&
H^I{}_Kd^{JK}+e^\rho b^Ib^J\cr}
\right]\ .
\end{equation}
In the last entry of $M$, $d^{IJ}$ is the inverse of $d_{IJ}$.  We
verify that
\begin{equation}
M^T=M,\qquad MLM^T=L^{-1}\ .
\end{equation}
This agrees with the bosonic action given in \cite{Senssd}.

\def\baselinestretch{1}
\normalsize
\newpage


\begin{thebibliography}{10}

\bibitem{Green}
M.~B. Green, J.~H. Schwarz and E.~Witten,
\newblock {\em Superstring Theory} (Cambridge University Press, 1987).

\bibitem{Pope}
M.~J. Duff, B.~E.~W. Nilsson and C.~N. Pope,
\newblock {\sl {K}aluza-{K}lein supergravity},
\newblock Phys. Rep. 130 (1986) 1.

\bibitem{Witten}
E.~Witten,
\newblock {\sl String theory dynamics in various dimensions},
\newblock IASSNS-HEP-95-18,\hfil\break
{\tt hep-th/9503124} (March 1995).

\bibitem{Luloop}
M.~J. Duff and J.~X. Lu,
\newblock {\sl Loop expansions and string/five-brane duality},
\newblock Nucl. Phys. B 357 (1991) 534.

\bibitem{Khurifour}
M.~J. Duff and R.~R. Khuri,
\newblock {\sl Four-dimensional string/string duality},
\newblock Nucl. Phys. B 411 (1994) 473.

\bibitem{Lublack}
M.~J. Duff and J.~X. Lu,
\newblock {\sl Black and super $p$-branes in diverse dimensions},
\newblock Nucl. Phys. B 416 (1994) 301.

\bibitem{Minasian}
M.~J. Duff and R.~Minasian,
\newblock {\sl Putting string/string duality to the test},
\newblock Nucl. Phys. B 436 (1995) 507.

\bibitem{Duffstrong}
M.~J. Duff,
\newblock {\sl Strong/weak coupling duality from the dual string},
\newblock Nucl. Phys. B 442 (1995) 47.

\bibitem{Khuristring}
M.~J. Duff, R.~R. Khuri and J.~X. Lu,
\newblock {\sl String solitons},
\newblock NI-94-017, CTP-TAMU-67/92, McGill/94-53, CERN-TH.7542/94,
{\tt hep-th/9412184}, to be published in Phys. Rep. (December 1994).

\bibitem{Duffclassical}
M.~J. Duff,
\newblock {\sl Classical/quantum duality},
\newblock in {\em Proceedings of the International High Energy Physics
  Conference, Glasgow} (June 1994).

\bibitem{Hull}
C.~M. Hull and P.~K. Townsend,
\newblock {\sl Unity of superstring dualities},
\newblock Nucl. Phys. B 438 (1995) 109.

\bibitem{Lustrings}
M.~J. Duff and J.~X. Lu,
\newblock {\sl Strings from fivebranes},
\newblock Phys. Rev. Lett. 66 (1991) 1402.

\bibitem{Lubetween}
M.~J. Duff and J.~X. Lu,
\newblock {\sl A duality between strings and fivebranes},
\newblock Class. Quantum Grav. 9 (1992) 1.

\bibitem{Dixonputting}
J.~A. Dixon, M.~J. Duff and J.~C. Plefka,
\newblock {\sl Putting string/five-brane duality to the test},
\newblock Phys. Rev. Lett. 69 (1992) 3009.

\bibitem{Izquierdo}
J.~M. Izquierdo and P.~K. Townsend,
\newblock {\sl Axionic defect anomalies and their cancellation},
\newblock Nucl. Phys. B 414 (1994) 93.

\bibitem{Blum}
J.~Blum and J.~A. Harvey,
\newblock {\sl Anomaly inflow for gauge defects},
\newblock Nucl. Phys. B 416 (1994) 119.

\bibitem{Vafa}
C.~Vafa and E.~Witten,
\newblock {\sl A one-loop test of string duality},
\newblock HUTP-95-A015, IASSNS-HEP-95-33, {\tt hep-th/9505053} (May 1995).

\bibitem{Dixoncoupling}
J.~A. Dixon, M.~J. Duff and E.~Sezgin,
\newblock {\sl The coupling of {Y}ang-{M}ills to extended objects},
\newblock Phys. Lett. B 279 (1992) 265.

\bibitem{Dixonchern}
J.~A. Dixon and M.~J. Duff,
\newblock {\sl {C}hern-{S}imons forms, {M}ickelsson-{F}adeev algebras and the
  $p$-branes},
\newblock Phys. Lett. B 296 (1992) 28.

\bibitem{Percacci}
R.~Percacci and E.~Sezgin,
\newblock {\sl Symmetries of $p$-branes},
\newblock Int. J. Mod. Phys. A 8 (1993) 5367.

\bibitem{Bergshoeff1}
E.~Bergshoeff, R.~Percacci, E.~Sezgin, K.~S. Stelle and P.~K. Townsend,
\newblock {\sl ${U}(1)$ extended gauge algebras in $p$-loop space},
\newblock Nucl. Phys. B 398 (1993) 343.

\bibitem{Cederwall}
M.~Cederwall, G.~Ferretti, B.~E.~W. Nilsson and A.~Westerberg,
\newblock {\sl Higher dimensional loop algebras, nonabelian extensions and
  $p$-branes},
\newblock Nucl. Phys. B 424 (1994) 97.

\bibitem{Duffsuper}
M.~J. Duff,
\newblock {\sl Supermembranes: The first fifteen weeks},
\newblock Class. Quantum Grav. 5 (1988) 189.

\bibitem{Townsendreview}
P.~K. Townsend,
\newblock {\sl Three lectures on supersymmetry and extended objects},
\newblock in {\em Proceedings of the 13th GIFT Seminar on Theoretical Physics:
  {\it Recent problems in Mathematical Physics}, Salamanca, Spain} (15-27 June
  1992).

\bibitem{Nilsson1}
M.~J. Duff and B.~E.~W. Nilsson,
\newblock {\sl Four-dimensional string theory from the {K3} lattice},
\newblock Phys. Lett. B 175 (1986) 417.

\bibitem{Nilsson2}
M.~J. Duff, B.~E.~W. Nilsson and C.~N. Pope,
\newblock {\sl Compactification of {$D = 11$} supergravity on {$K3 \times
  T^3$}},
\newblock Phys. Lett. B 129 (1983) 39.

\bibitem{Narain}
K.~S. Narain,
\newblock {\sl New heterotic string theories in uncompactified dimensions $<
  10$},
\newblock Phys. Lett. B 169 (1986) 41.

\bibitem{Narain2}
K.~S. Narain, M.~H. Sarmadi and E.~Witten,
\newblock {\sl A note on toroidal compactification of heterotic string theory},
\newblock Nucl. Phys. B 279 (1987) 369.

\bibitem{Seiberg}
N.~Seiberg,
\newblock {\sl Observations on the moduli space of superconformal field
  theories},
\newblock Nucl. Phys. B 303 (1988) 286.

\bibitem{Aspinwall1}
P.~S. Aspinwall and D.~R. Morrison,
\newblock {\sl String theory on {K3} surfaces},
\newblock DUK-TH-94-68, IASSNS-HEP-94/23, {\tt hep-th/9404151} (April 1994).

\bibitem{Bergshoeff2}
E.~Bergshoeff, E.~Sezgin and P.~K. Townsend,
\newblock {\sl Supermembranes and eleven-dimensional supergravity},
\newblock Phys. Lett. B 189 (1987) 75.

\bibitem{Bergshoeff-annp}
E.~Bergshoeff, E.~Sezgin and P.~K. Townsend,
\newblock {\sl Properties of the eleven-dimensional supermembrane theory},
\newblock Ann. Phys. 185 (1988) 330.

\bibitem{Howe}
M.~J. Duff, P.~S. Howe, T.~Inami and K.~S. Stelle,
\newblock {\sl Superstrings in {$D = 10$} from supermembranes in {$D = 11$}},
\newblock Phys. Lett. B 191 (1987) 70.

\bibitem{Dufflu}
M.~J. Duff and J.~X. Lu,
\newblock {\sl Duality for strings and membranes},
\newblock in Duff, Sezgin and Pope, editors, {\em Supermembranes and Physics in
  $2+1$ Dimensions} (World Scientific, 1990);
\newblock also published in Arnowitt, Bryan, Duff, Nanopoulos, Pope and Sezgin,
  editors, {\it Strings 90} (World Scientific, 1991).

\bibitem{Luduality}
M.~J. Duff and J.~X. Lu,
\newblock {\sl Duality rotations in membrane theory},
\newblock Nucl. Phys. B 347 (1990) 394.

\bibitem{Scherk}
E.~Cremmer, J.~Scherk and S.~Ferrara,
\newblock {\sl {$SU(4)$} invariant supergravity theory},
\newblock Phys. Lett. B 74 (1978) 61.

\bibitem{Cremmer}
E.~Cremmer and B.~Julia,
\newblock {\sl The {$SO(8)$} supergravity},
\newblock Nucl. Phys. B 159 (1979) 141.

\bibitem{Marcus}
N.~Marcus and J.~H. Schwarz,
\newblock {\sl Three-dimensional supergravity theories},
\newblock Nucl. Phys. B 228 (1983) 145.

\bibitem{Dufffradkin}
M.~J. Duff,
\newblock {\sl {$E_8\times SO(16)$} symmetry of {$D=11$} supergravity?},
\newblock in Batalin, Isham and Vilkovisky, editors, {\em Quantum Field Theory
  and Quantum Statistics, Essays in Honour of E. S. Fradkin} (Adam Hilger,
  1986).

\bibitem{Giveonreview}
A.~Giveon, M.~Porrati and E.~Rabinovici,
\newblock {\sl Target space duality in string theory},
\newblock Phys. Rep. 244 (1994) 77.

\bibitem{Font}
A.~Font, L.~Ibanez, D.~Lust and F.~Quevedo,
\newblock {\sl Strong-weak coupling duality and nonperturbative effects in
  string theory},
\newblock Phys. Lett. B 249 (1990) 35.

\bibitem{Rey}
S.-J. Rey,
\newblock {\sl The confining phase of superstrings and axionic strings},
\newblock Phys. Rev. D 43 (1991) 526.

\bibitem{Kalara}
S.~Kalara and D.~V. Nanopoulos,
\newblock {\sl String duality and black holes},
\newblock Phys. Lett. B 267 (1991) 343.

\bibitem{Sen2}
A.~Sen,
\newblock {\sl Electric magnetic duality in string theory},
\newblock Nucl. Phys. B 404 (1993) 109.

\bibitem{Sen3}
A.~Sen,
\newblock {\sl Quantization of dyon charge and electric magnetic duality in
  string theory},
\newblock Phys. Lett. B 303 (1993) 22.

\bibitem{Schwarz1}
J.~H. Schwarz and A.~Sen,
\newblock {\sl Duality symmetries of 4-{$D$} heterotic strings},
\newblock Phys. Lett. B 312 (1993) 105.

\bibitem{Schwarz2}
J.~H. Schwarz and A.~Sen,
\newblock {\sl Duality symmetric actions},
\newblock Nucl. Phys. B 411 (1994) 35.

\bibitem{Binetruy}
P.~Binetruy,
\newblock {\sl Dilaton, moduli and string/five-brane duality as seen from
  four-dimensions},
\newblock Phys. Lett. B 315 (1993) 80.

\bibitem{Sen4}
A.~Sen,
\newblock {\sl {$SL(2,Z)$} duality and magnetically charged strings},
\newblock Mod. Phys. Lett. A 8 (1993) 5079.

\bibitem{Rahmfeld}
M.~J. Duff and J.~Rahmfeld,
\newblock {\sl Massive string states as extreme black holes},
\newblock Phys. Lett. B 345 (1995) 441.

\bibitem{Gauntlett}
J.~P. Gauntlett and J.~A. Harvey,
\newblock {\sl {S} duality and the spectrum of magnetic monopoles in heterotic
  string theory},
\newblock EFI-94-11, {\tt hep-th/9407111} (July 1994).

\bibitem{Sen7}
A.~Sen,
\newblock {\sl Strong-weak coupling duality in three-dimensional string
  theory},
\newblock Nucl. Phys. B 434 (1995) 179.

\bibitem{Rahmfeld2}
M.~J. Duff, S.~Ferrara, R.~R. Khuri and J.~Rahmfeld,
\newblock {\sl Supersymmetry and dual string solitons},
\newblock NI-94-034, CTP-TAMU-50/94, {\tt hep-th/9506057} (June 1995).

\bibitem{Stelle}
M.~J. Duff and K.~S. Stelle,
\newblock {\sl Multimembrane solutions of {$D = 11$} supergravity},
\newblock Phys. Lett. B 253 (1991) 113.

\bibitem{Dabholkar}
A.~Dabholkar, G.~Gibbons, J.~A. Harvey and F.~Ruiz-Ruiz,
\newblock {\sl Superstrings and solitons},
\newblock Nucl. Phys. B 340 (1990) 33.

\bibitem{Luelem}
M.~J. Duff and J.~X. Lu,
\newblock {\sl Elementary fivebrane solutions of {$D = 10$} supergravity},
\newblock Nucl. Phys. B 354 (1991) 141.

\bibitem{Callan1}
C.~G. Callan, J.~A. Harvey and A.~Strominger,
\newblock {\sl World sheet approach to heterotic instantons and solitons},
\newblock Nucl. Phys. B 359 (1991) 611.

\bibitem{Callan2}
C.~G. Callan, J.~A. Harvey and A.~Strominger,
\newblock {\sl Worldbrane actions for string solitons},
\newblock Nucl. Phys. B 367 (1991) 60.

\bibitem{Strominger}
A.~Strominger,
\newblock {\sl Heterotic solitons},
\newblock Nucl. Phys. B 343 (1990) 167;
\newblock E: Nucl. Phys. B 353 (1991) 565.

\bibitem{Horowitz1}
G.~T. Horowitz and A.~Strominger,
\newblock {\sl Black strings and $p$-branes},
\newblock Nucl. Phys. B 360 (1991) 197.

\bibitem{Luscan}
M.~J. Duff and J.~X. Lu,
\newblock {\sl Type {II} $p$-branes: The brane scan revisited},
\newblock Nucl. Phys. B 390 (1993) 276.

\bibitem{Guven}
R.~Guven,
\newblock {\sl Black $p$-brane solutions of $d=11$ supergravity theory},
\newblock Phys. Lett. B 276 (1992) 49.

\bibitem{Gibbonstownsend}
G.~W. Gibbons and P.~K. Townsend,
\newblock {\sl Vacuum interpolation in supergravity via super $p$-branes},
\newblock Phys. Rev. Lett. 71 (1993) 3754.

\bibitem{Townsendeleven}
P.~K. Townsend,
\newblock {\sl The eleven-dimensional supermembrane revisited},
\newblock {\tt hep-th/9501068} (1995).

\bibitem{Gibbons}
M.~J. Duff, G.~W. Gibbons and P.~K. Townsend,
\newblock {\sl Macroscopic superstrings as interpolating solitons},
\newblock Phys. Lett. B 332 (1994) 321.

\bibitem{Horowitz2}
G.~W. Gibbons, G.~T. Horowitz and P.~K. Townsend,
\newblock {\sl Higher dimensional resolution of dilatonic black hole
  singularities},
\newblock Class. Quantum Grav. 12 (1995) 297.

\bibitem{Khurinew}
M.~J. Duff, R.~R. Khuri, R.~Minasian and J.~Rahmfeld,
\newblock {\sl New black hole, string and membrane solutions of the four
  dimensional heterotic string},
\newblock Nucl. Phys. B 418 (1994) 195.

\bibitem{Townsendseven}
P.~K. Townsend,
\newblock {\sl String-membrane duality in seven dimensions},
\newblock DAMTP-R/95/12, {\tt hep-th/9504095} (April 1995).

\bibitem{Harvey}
J.~A. Harvey and A.~Strominger,
\newblock {\sl The heterotic string is a soliton},
\newblock EFI-95-16, {\tt hep-th/9504047} (April 1995).

\bibitem{Bars}
I.~Bars,
\newblock {\sl Stringy evidence for ${D}=11$ structure in strongly coupled
  {Type} {II-A} superstring},
\newblock USC-95/HEP-B2, {\tt hep-th/9503228} (March 1995).

\bibitem{Aspinwall2}
P.~S. Aspinwall and D.~R. Morrison,
\newblock {\sl {U}-duality and integral structures},
\newblock CLNS-95/1334, {\tt hep-th/9505025} (May 1995).

\bibitem{Alvarez}
L.~Alvarez-Gaum\'e and E.~Witten,
\newblock {\sl Gravitational anomalies},
\newblock Nucl. Phys. B 234 (1984) 269.

\bibitem{Ginsparg}
L.~Alvarez-Gaum\'e and P.~Ginsparg,
\newblock {\sl The structure of gauge and gravitational anomalies},
\newblock Ann. Phys. 161 (1985) 423.

\bibitem{Lerche}
W.~Lerche, B.~E.~W. Nilsson and A.~N. Schellekens,
\newblock {\sl Heterotic string loop calculation of the anomaly cancelling
  term},
\newblock Nucl. Phys. B 289 (1987) 609.

\bibitem{lnsw}
W.~Lerche, B.~E.~W. Nilsson, A.~N. Schellekens and N.~P. Warner,
\newblock {\sl Anomaly cancelling terms from the elliptic genus},
\newblock Nucl. Phys. B 299 (1988) 91.

\bibitem{swanomlet}
A.~N. Schellekens and N.~P. Warner,
\newblock {\sl Anomalies and modular invariance in string theory},
\newblock Phys. Lett. B 177 (1986) 317.

\bibitem{swanomnp}
A.~N. Schellekens and N.~P. Warner,
\newblock {\sl Anomalies, characters and strings},
\newblock Nucl. Phys. B 287 (1987) 317.

\bibitem{Luremarks}
M.~J. Duff and J.~X. Lu,
\newblock {\sl Remarks on string/fivebrane duality},
\newblock Nucl. Phys. B 354 (1991) 129.

\bibitem{Gross}
D.~J. Gross, J.~A. Harvey, E.~Martinec and R.~Rohm,
\newblock {\sl Heterotic string theory. 1. the free heterotic string},
\newblock Nucl. Phys. B 256 (1985) 253.

\bibitem{Hullwitten}
C.~M. Hull and E.~Witten,
\newblock {\sl Supersymmetric sigma models and the heterotic string},
\newblock Phys. Lett. B 160 (1985) 398.

\bibitem{Senssd}
A.~Sen,
\newblock {\sl String string duality conjecture in six dimensions and charged
  solitonic strings},
\newblock TIFR-TH-95-16, {\tt hep-th/9504027} (April 1995).

\bibitem{Percacci2}
R.~Percacci and E.~Sezgin,
\newblock {\sl On target space duality in $p$-branes},
\newblock Mod. Phys. Lett. A 10 (1995) 441.

\bibitem{Luthree}
M.~J. Duff and J.~X. Lu,
\newblock {\sl The selfdual {T}ype {IIB} superthreebrane},
\newblock Phys. Lett. B 273 (1991) 409.

\bibitem{Wittendual}
E.~Witten,
\newblock {\sl On {S}-duality in abelian gauge theory},
\newblock IASSNS-HEP-95-36, {\tt hep-th/9505186} (May 1995).

\bibitem{Verlinde}
E.~Verlinde,
\newblock {\sl Global aspects of electric-magnetic duality},
\newblock CERN-TH/95-146,\hfil\break
{\tt hep-th/9506011} (May 1995).

\bibitem{Gibbonsnew}
G.~W. Gibbons and D.~A. Rasheed,
\newblock {\sl Electric-magnetic duality rotations in non-linear
  electrodynamics},
\newblock DAMTP preprint (June 1995).

\bibitem{Dyson}
F.~J. Dyson,
\newblock {\em Infinite in All Directions: Gifford lectures given at Aberdeen,
  Scotland} (Harper \& Row, New York, 1988).

\end{thebibliography}



\newpage

\begin{figure}
\figlab{a}
\epsfysize=3.3in
\centerline{\epsfbox{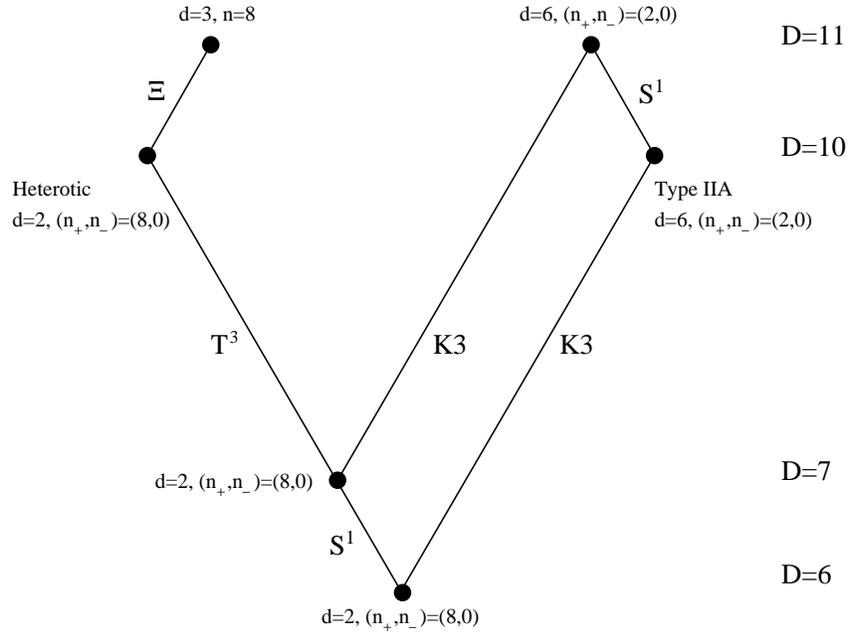}}
\vskip.7in
\figlab{b}
\epsfysize=3.3in
\centerline{\epsfbox{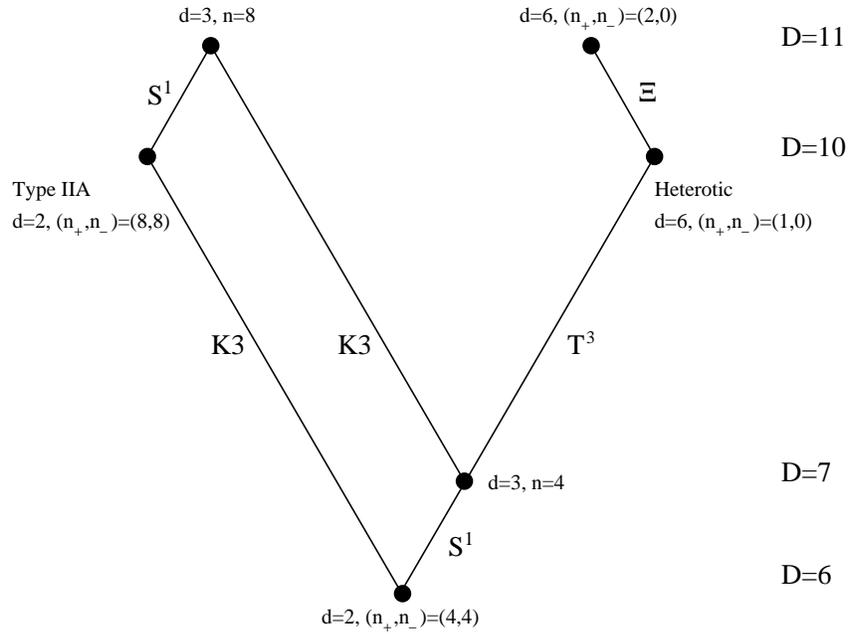}}
\vskip.3in
\caption{Compactifications relating $(a)$ the Type $IIA$ fivebrane to the
heterotic string and $(b)$ the heterotic fivebrane to the Type $IIA$ string.
Worldvolume supersymmetries are indicated.}
\label{fig:one}
\end{figure}

\newpage
\begin{figure}
\figlab{a}
\epsfysize=3.4in
\centerline{\epsfbox{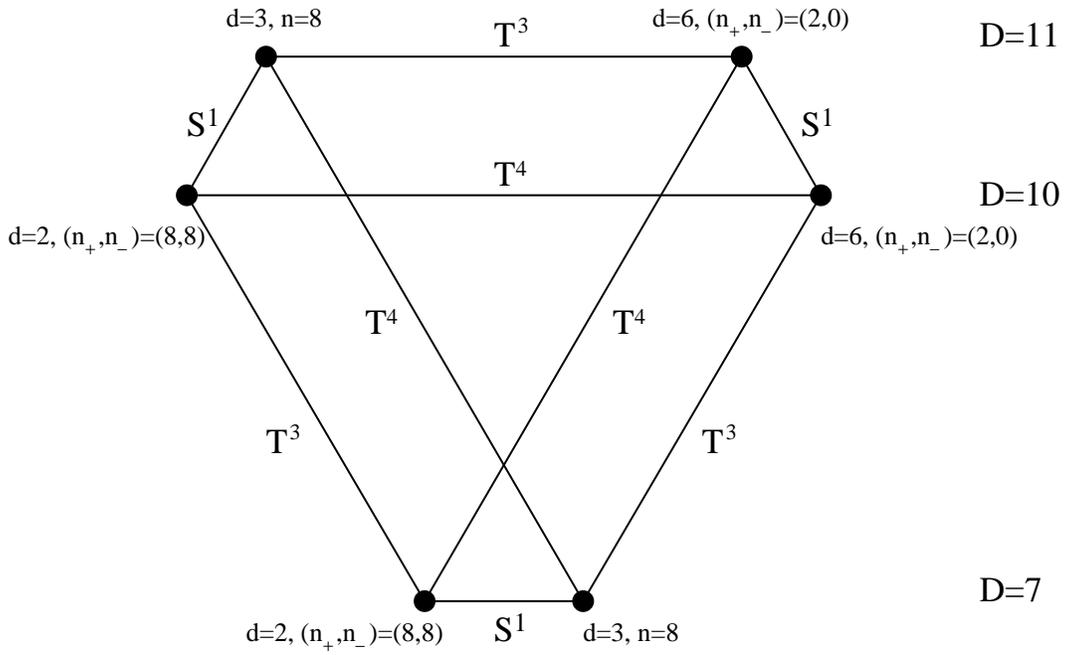}}
\vskip.7in
\figlab{b}
\epsfysize=3.4in
\centerline{\epsfbox{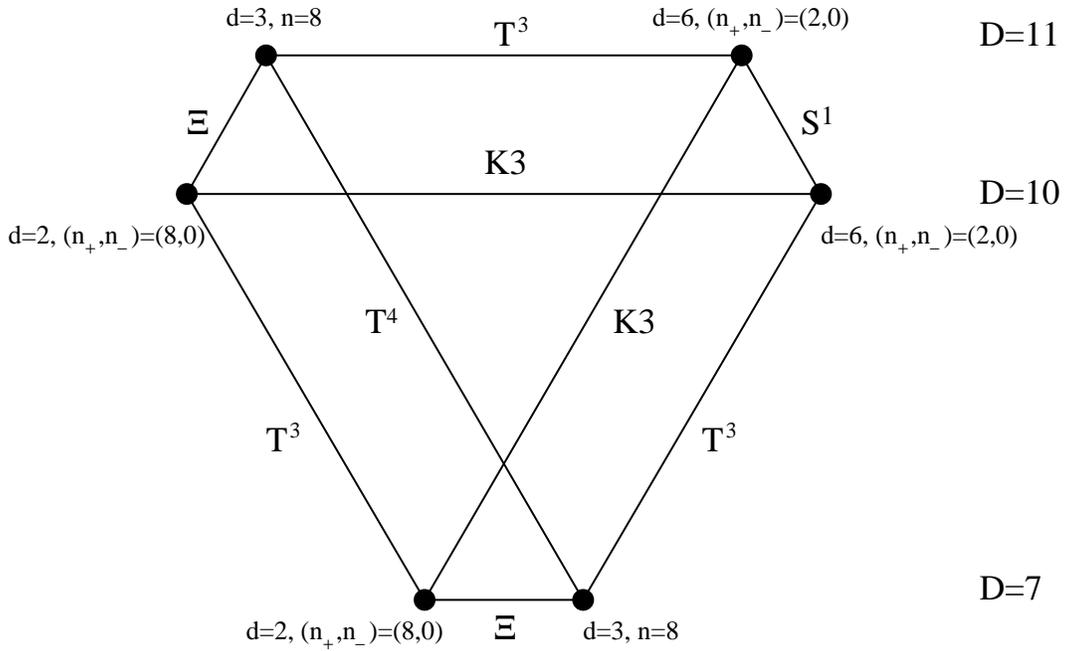}}
\vskip.6in
\caption{Compactifications incorporating worldvolume reductions.}
\label{fig:two}
\end{figure}

\newpage
\begin{figure}
\centerline{\epsfbox{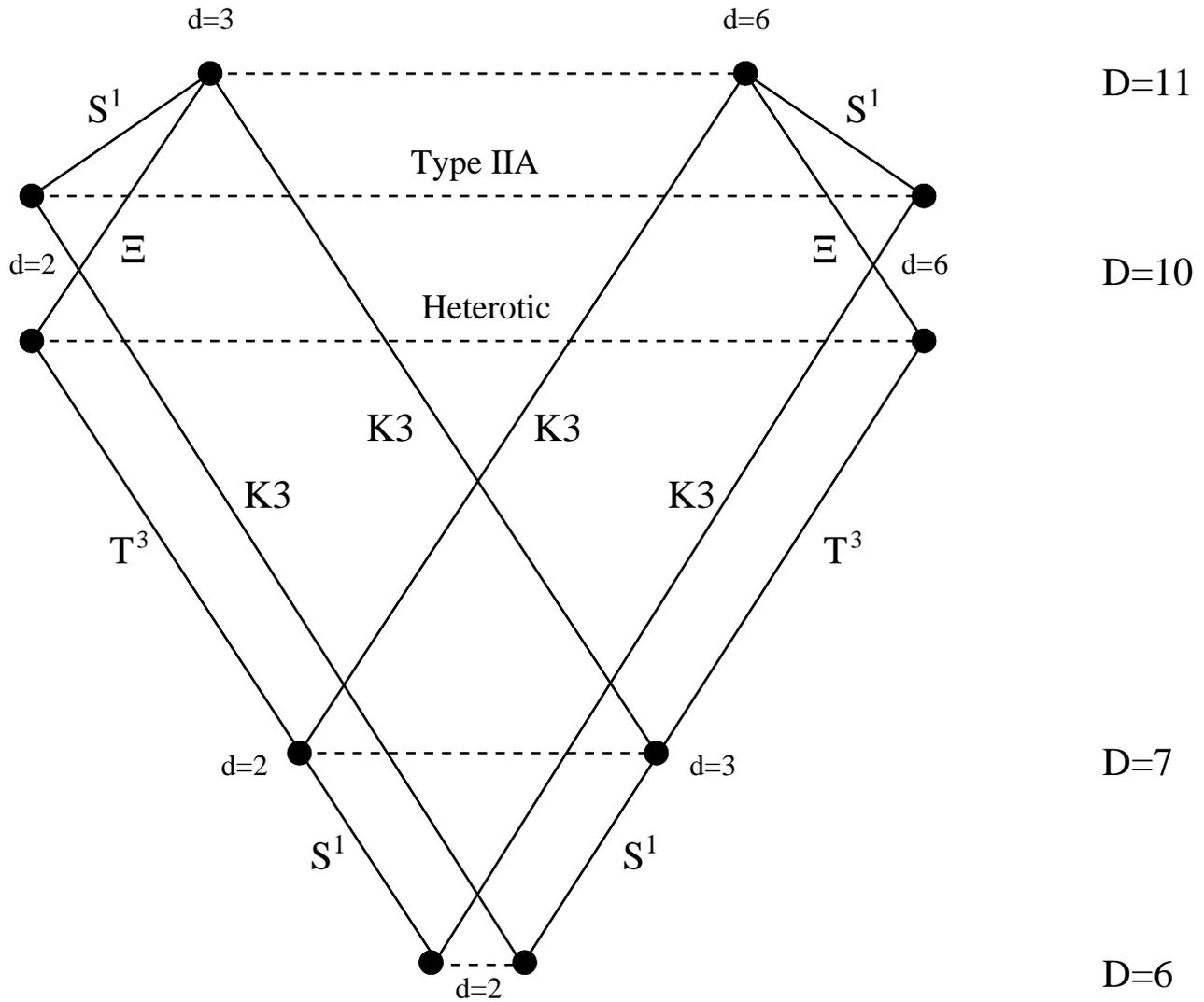}}
\vskip.8in
\caption{A superposition of Figs.~1 $(a)$ and $(b)$, illustrating
strong/weak coupling dualities (denoted by the dotted lines).}
\label{fig:three}
\end{figure}

\end{document}